%% file: sdd.tex
\documentclass[letterpaper,11pt]{report}

\pdfoutput=1

\usepackage{graphicx}
\usepackage{latexsym}
\usepackage{makeidx}
\usepackage{url}
\usepackage{hyperref}

\makeindex

\input{commands}

\begin{document}
\pagestyle{headings}
%---------------------- Title Page Start --------------------
%\begin{titlepage}  %alternative method for specifying
%\end{titlepage}

\title{Ftklipse -- Design and Implementation of an Extendable Computer Forensics Environment\\
Specification Design Document}
\author{Marc-Andr\'e Laverdi\`ere \and Serguei A. Mokhov \and Suhasini Tsapa \and Djamel Benredjem}
\date{April 24, 2006}
\maketitle
%---------------------- Title Page End --------------------

\include{toc}
\include{introduction}
\include{system-overview}
\include{detailed-design}
\include{conclusion}
\include{references}
\include{appendix}

%appendices go here
%\include{appendix}

\printindex

\end{document}

%% file: commands.tex
\newcommand{\xf}[1]{Figure~\ref{#1}}

\newcommand{\xt}[1]{Table~\ref{#1}}

\newcommand{\java}{{Java\index{Java}}}

\newcommand{\file}[1]{\texttt{#1}\index{Files!#1}}
\newcommand{\tool}[1]{\texttt{#1}\index{Tools!#1}}

\newcommand{\api}[1]{\texttt{#1}\index{API!#1}}

\newcommand{\marf}[0]{MARF\index{Tools!MARF}\index{Libraries!MARF}}

\newcommand{\lucidL}[1]{{$\mathit{Lucid}$}($L$) }

		{}

\def\myvert{\raise 2.27pt \hbox{\vrule depth 0pt height 8pt width 0.2mm}}
\def\myarrow{\hspace*{0.43mm}%
             \raise 2.29pt\hbox{\vrule depth 0pt height 8pt width 0.16mm}%
             \hspace*{-0.32mm}%
             $\longrightarrow$
             \ %
             }

%% file: toc.tex
\pagenumbering{roman}
\tableofcontents
\clearpage
\pagenumbering{arabic}

%\listoffigures
%\listoftables

%% file: introduction.tex
%---------------------- Introduction Start ------------------
\chapter{Introduction}
\index{Introduction}

This chapter briefly presents the purpose and the scope of the work on the Ftklipse project
with a subset of relevant definitions and acronyms. All these aspects
are detailed to some extent later through the document.

\section{Purpose}
\index{Introduction!Purpose}

To design and implement a plugin-based environment that allows to integrate forensic tools working together
to support programming tasks and addition of new tools.
Integration is done through GUI components.

\section{Scope}
\index{Introduction!Scope}

The end-product enviroment must have user friendly GUI, configuration capabilities, plug-in capabilities to insert/inject new tools,
case management, and chain of custody capabilities, along with evidence gathering capabilities,
evidence preservation capabilities, and, finally report generation capabilities.
A subset of these requirements has been implemented in Ftklipse, which is detailed
throughout the rest of this document.

\section{Definitions and Acronyms}
\index{Introduction!Definitions and Acronyms}

\begin{description}

\item[Cryptographic Hash Function] Function mapping input data of an arbritary size to  a fixed-sized output that is highly collision resistant.

\item[Digital evidence] Information stored or transmitted in binary form that may be relied upon in court.

\item[\tool{dcfldd}] Enhanced DD imager with built-in hashing, works like \tool{dd} from command line.
\begin{description}
	\item[Hashing on-the-fly] \tool{dcfldd} can hash the input data as it is being transferred, helping to ensure data integrity.
	\item[Status output] \tool{dcfldd} can update the user of its progress in terms of the amount of data transferred and how much longer operation will take.
	\item[Flexible disk wipes] \tool{dcfldd} can be used to wipe disks quickly and with a known pattern if desired.
	\item[Image/wipe verify] \tool{dcfldd} can verify that a target drive is a bit-for-bit match of the specified input file or pattern.
	\item[Multiple outputs] \tool{dcfldd} can output to multiple files or disks at the same time.
	\item[Split output] \tool{dcfldd} can split output to multiple files with more configurability than the split command.
	\item[Piped output and logs] \tool{dcfldd} can send all its log data and output to commands as well as files natively.
\end{description}

\item[Documentation] Written notes, audio/videotapes, printed forms, sketches, and/or photographs that form a detailed record of the
scene, evidence recovered, and actions taken during the search of the scene.

\item[JVM] The Java Virtual Machine. Program and framework allowing the execution of program
	developped using the Java programming language.

\item[Magnetic media] A disk, tape, cartridge, diskette, or cassette that is used to store data magnetically.

\item[Steganography] It simply takes one piece of information and hides it within another. Computer files (images, sounds recordings,
even disks) contain unused or insignificant areas of data. Steganography takes advantage of these areas, replacing them with
information (encrypted mail, for instance). The files can then be exchanged without anyone knowing what really lies inside of them.
For example, an image of the space shuttle landing might contain a private letter to a friend. A recording of a short sentence might contain your
company's plans for a secret new product.
Steganography can also be used to place a hidden ``trademark'' in images, music, and software, a technique referred to as
watermarking.

\item[SWT] The Standard Widget Toolkit \cite{swt}, a set of graphical user interface components provided by the Eclipse framework.

\item[Temporary and swap files] Many computers use operating systems and applications that store data temporarily on the hard
drive. These files, which are generally hidden and inaccessible,may contain information that the investigator finds useful.

\end{description}

%---------------------- Introduction End --------------------

% EOF

%% file: system-overview.tex
\chapter{System Overview}
\index{System Overview}

In this chapter, we examine the architecture of our implementation of Ftklipse. We first introduce our
architectural philosophy before informing
the reader about the Siemens Four View Model, an architectural methodology for the conception of large-scale software systems.
Afterwards, we examine each of the view, as architected for our system.
Finally, we conclude with other software engineering matters that were of high importance
in the development of our implementation.

\section{Architectural Strategies}
\index{System Overview!Architectural Strategies}

Our principles are:

\begin{description}

\item [Platform independence] We target systems that are capabale of running a JVM.

\item [The Eclipse plug-in based environment] slightly imitating the  MVC (Model-view -Controller) pattern,
to map the traditional input, processing, output roles into the GUI realm.
In Eclipse model, a plug-in may be related to another plug-in by one of two relationships:

\begin{description}
\item [Dependency] The roles in this relationship are dependent plug-in and prerequsite plug-in. A prerequisite plug-in
supports the function of a dependent plug-in.

\item
[Extension] The roles in this relationship are host plug-in and extender plug-in.  An extender plug-in extends
the functions of a host plug -in.
\end{description}

\item [Database independent API] will allows us to swap database engines on-the-fly.

\item [Reasonable Efficiency] We will architect and implement an efficient system, but will avoid advanced programming
tricks that improve the efficiency at the cost of maintainability and readability.
\item [Simplicity And Maintainability]  We will target a simplistic and easy to maintain organization of the source.
\item [Architectural Consistency] We will consistently implement our architectural approach.
\item [Separation of Concerns] We will isolate separate concerns between modules and within modules to encourage reuse and code simplicity.
\end{description}

\section{System Architecture}

\subsection{Module View}

\subsubsection{Layering}

We divided our application between layers.
The top level has a front-end and a back-end.
The frontend comprised a collection of
GUI modules provided by and customized from eclipse
as well as custom-designd by the team.
The backend consists of supporting functionality
and specifically database management, report generation,
and external tool invocation.

\subsubsection{Interface Design}

Several interfaces had to be designed for the architecture to work
All the backend modules
have an interface they expose to the frontend to use.
Thus, there are interfaces between, GUI-to-External-Tools,
GUI-to-Database, and GUI-to-Report-Generation. All these are
designed to be swappable and highly modular so any component
series can be replaced at any time with little or no change
to the code. The interfaces (\api{FtklipseCommonDatabaseBridge} and \api{IDatabaseAdapter},
\api{ITool} and \api{IToolExecutor}, and \api{IReportGenerator} and \api{ReportGeneratorFactory}) are presented in the detiled
design chapter.

\section{Execution View}

\subsection{Runtime Entities}

In the case of our application, there is hosting run-time environment that of
Eclipse. The application can run within Eclipse IDE or be a stand-alone with
a minimal subset of the Eclipse run-time. By nature, a JVM machine is executing
all the environment and all GUI-based applications are multi-threaded to avoid
blockage on user's input. Additionally, depending on the database engine used
behind the scenes, it may as well be multi-threaded to provide concurrent access
and connection pooling.

\subsection{Communication Paths}

It was resolved that the modules would all communicate through message passing
between methods. Communication to the database depends on the database adapter,
and in our sample implementation is done through and in-process JDBC driver.
Additionally, Java's reflection is used to discover instantiation communication
paths at run-time for pluggable modules.

\subsection{Execution Configuration}

Execution configuration in Ftklipse has to do with where its
\file{data} directory is. The \file{data} directory is always
local to where the application was ran from. The directory
contains the main case database in the \file{ftklipsedb.*}
files as well as numerical directorys with case ID with
imported evidence files. Additionall configuration for application
is located in \file{plugin.properties} and \file{plugin.xml} files.

\section{Coding Standards and Project Management}
\label{sect:standards}

In order to produce high-quality code, we decided to normalize on the OpenBSD style.
We also decided to use \tool{javadoc} source code documentation style for
its completeness and the automated tool support.
We used Subversion (\tool{svn}) \cite{svn} in order to manage the source code, makefile,
and documentation revisions provided by \url{SourceForge.net}.

%\section{Software Interface Design}

%\subsection{User Interface}

%\subsection{User Interface Prototypes}

%{\todo}

%\subsubsection{Specific Tools}

%{\todo}

%\subsubsection{Batch Operations}

%{\todo}

%\subsubsection{Generic Tool}

%{\todo}

%\subsection{Software Interface}

%\subsection{Hardware Interface}

% EOF

%% file: detailed-design.tex
\chapter{Detailed System Design}
\index{Detailed System Design}
\index{Design}

\begin{itemize}
\item Case management: Investigations are organized by cases, which can contain one or more evidences.
Each evidence can contain one or more file system images to analyze;

\item Evidence Gathering using integrated and plug-in tools;

\item Evidence Integrity validation using a hash function;

\item Evidence Import from any media to an existing case;

\item Logging of all operations performed on the evidence;

\item Validation of integrity of evidence after each operation over it;

\item Display of evidence in read-only mode either in ASCII, Unicode or Hex formats;

\item Recording of investigative notes for each piece of evidence;

\item Capability to extract a part of the evidence into another file;

\item Capability to copy and rename the copy of the evidence;

\item Generation of reports in PDF and \LaTeX2e formats that includes listing of the evidence in the case,
a printout of selected parts of the evidence, the investigative notes related to selected parts of
the evidence and a customized executive summary, introduction, and conclusion. It also integrates
the chain of custodity information for each part of the evidence displaying the principal, timestamp
and operation performed on the evidence.

\item An extendable set of tools through a plug-in architecture;

\item General as well as tool-specific defaults and configuration screens;
\end{itemize}

\section{Class Diagrams}
\index{Design!Class Diagrams}

We have a number of class diagrams representing the majore modules
and their relationships. Please located the detailed descriprion
of the modules in the generated HTML of javadoc or the javadoc
comments themeselves in the \file{doc/javadoc} directory.

The basic UI classes are in \xf{fig:uml:ui}.
The prototype internal access control classes are in \xf{fig:uml:acl}.
The main database abstraction is in \xf{fig:uml:db}.
Next, concrete database adatpters are in \xf{fig:uml:dbadapters.}.
Further, the database- and UI-indepedent database objects
data structures are in \xf{fig:uml:dbobjects}. The report
generation-related API is in \xf{fig:uml:reporting}. Finally,
the external tools invocation framework is in \xf{fig:uml:tools}.

\begin{figure}
\includegraphics[width=\textwidth, keepaspectratio=true]{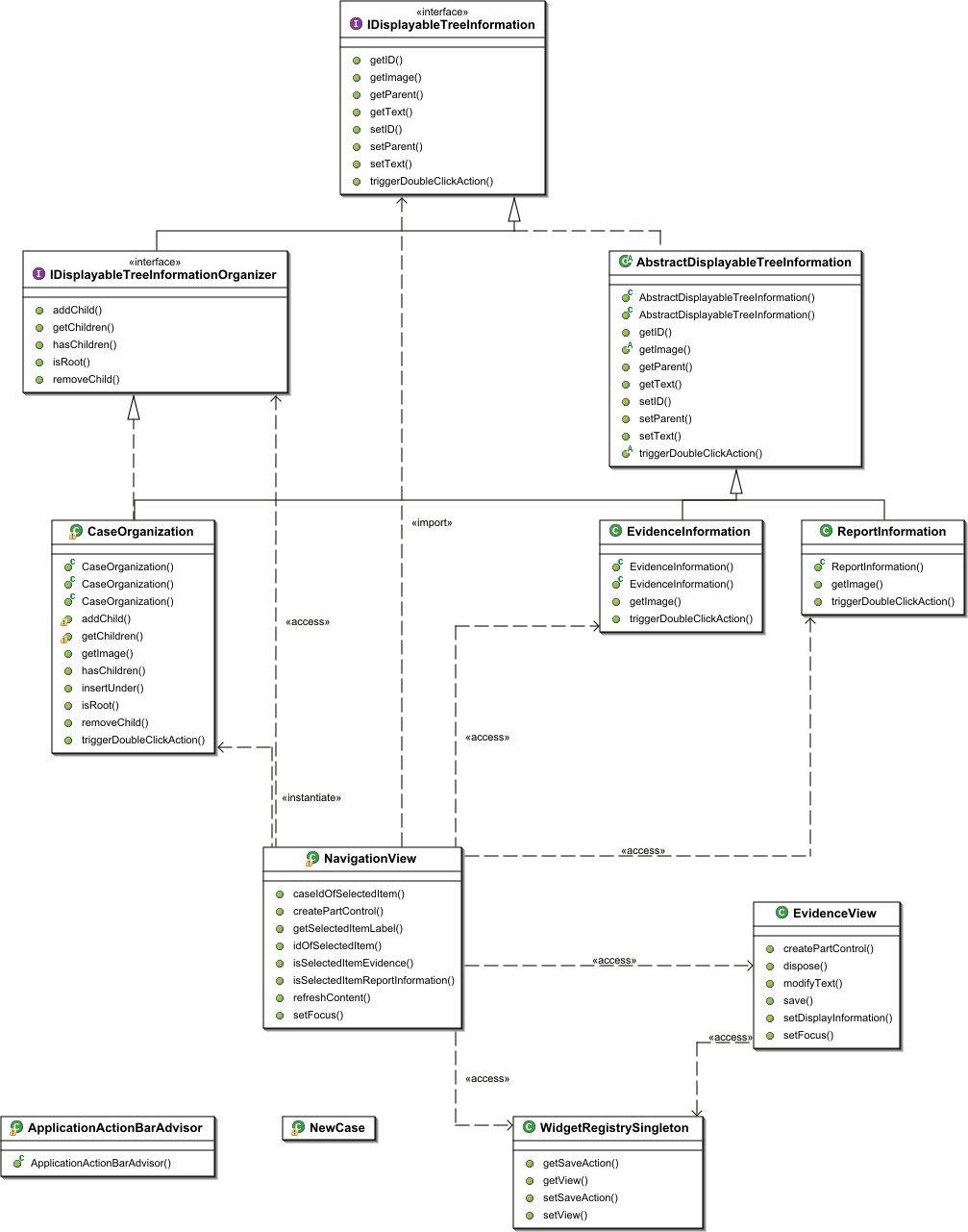}
\caption{Class Diagram for the basic User Interface}
\label{fig:uml:ui}
\end{figure}

\begin{figure}
\includegraphics[width=\textwidth, keepaspectratio=true]{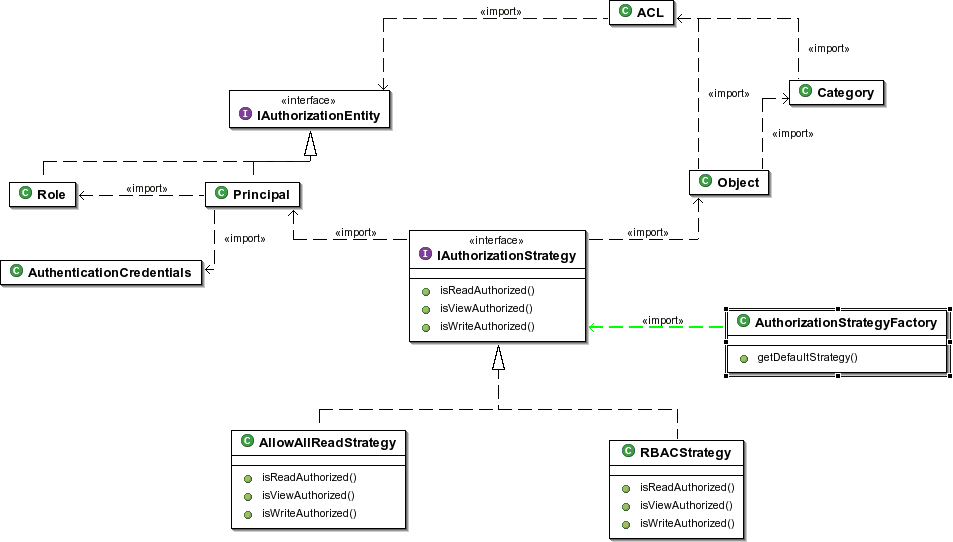}
\caption{Class Diagram for Access Control Framework}
\label{fig:uml:acl}
\end{figure}

\begin{figure}
\includegraphics[width=\textwidth, keepaspectratio=true]{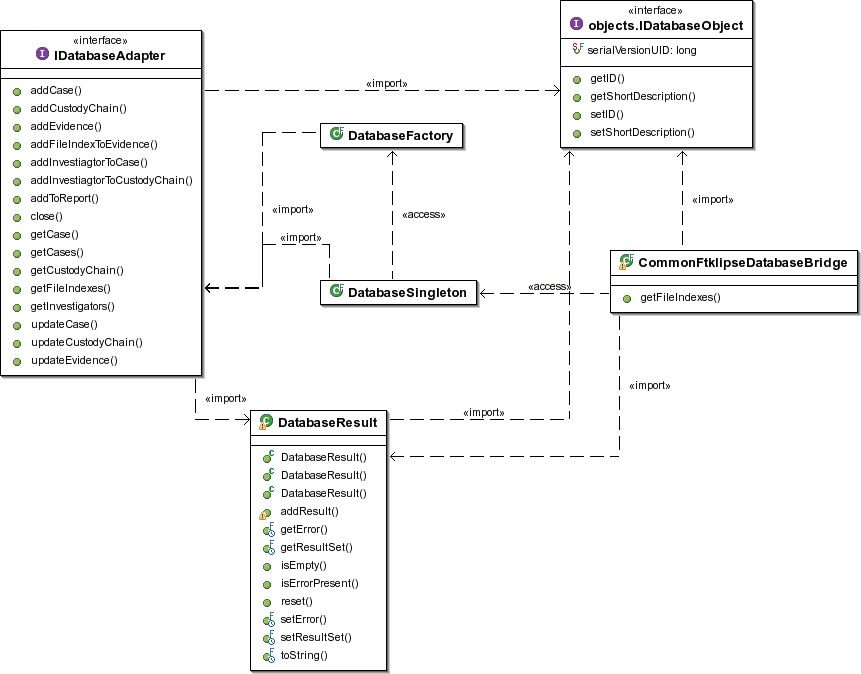}
\caption{Class Diagram for the Database Root Package}
\label{fig:uml:db}
\end{figure}

\begin{figure}
\includegraphics[width=\textwidth, keepaspectratio=true]{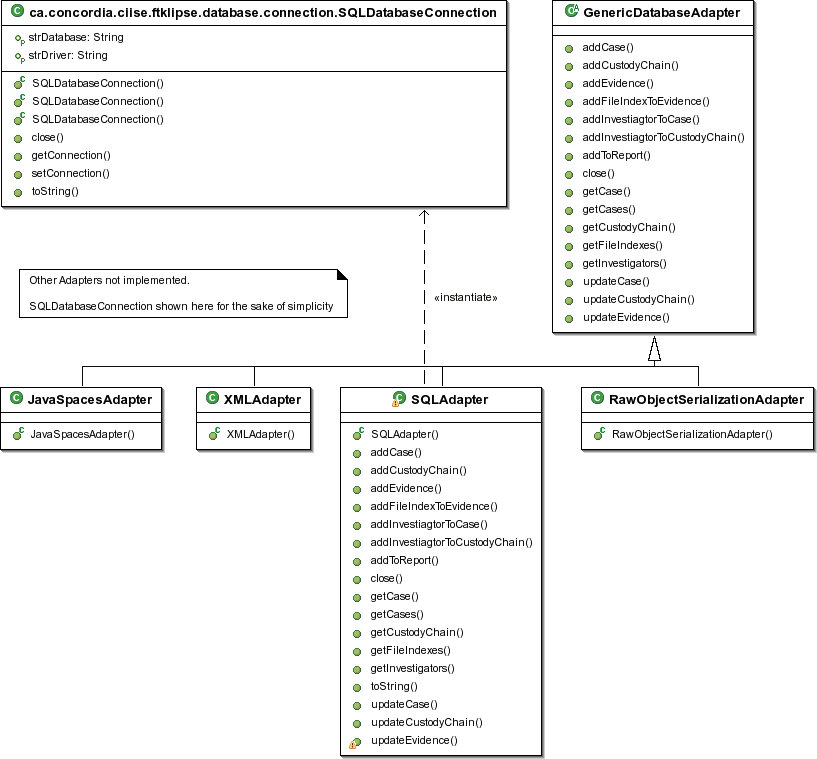}
\caption{Class Diagram for the Database Adapters}
\label{fig:uml:dbadapters}
\end{figure}

\begin{figure}
\includegraphics[width=\textwidth, keepaspectratio=true]{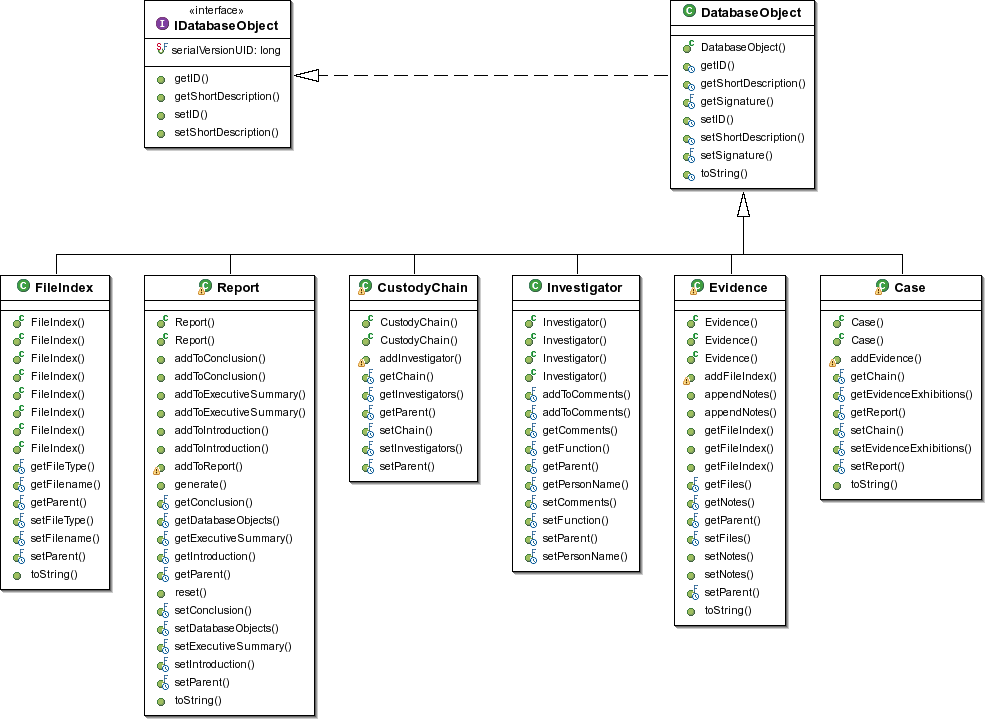}
\caption{Class Diagram for the Database Objects}
\label{fig:uml:dbobjects}
\end{figure}

\begin{figure}
\includegraphics[width=\textwidth, keepaspectratio=true]{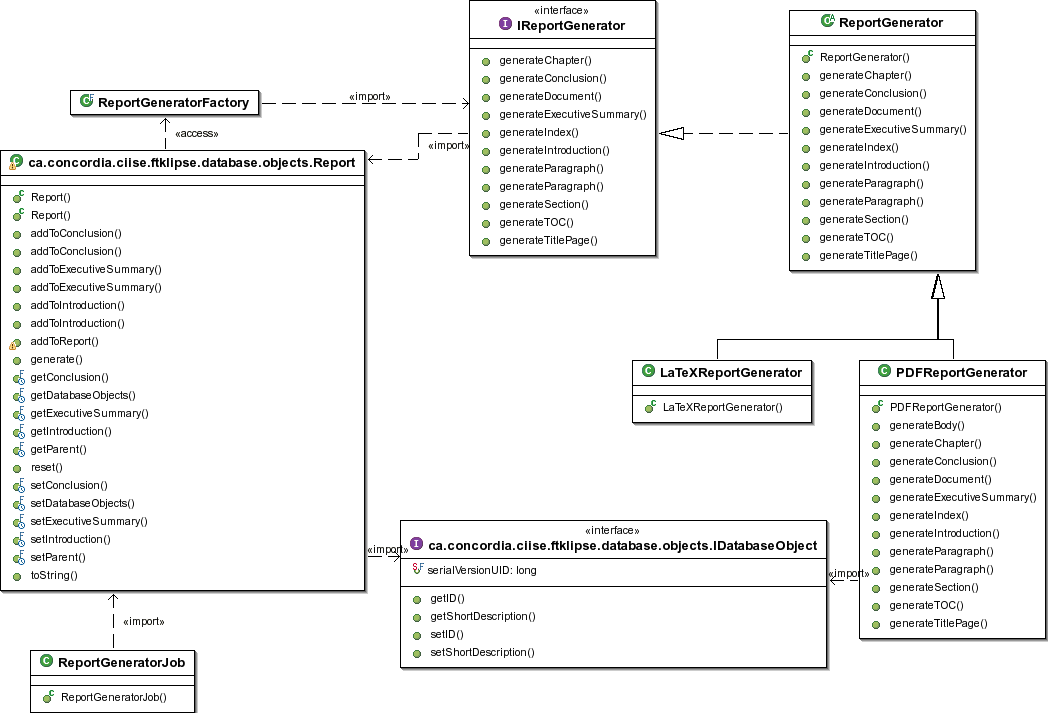}
\caption{Class Diagram for the Report Generation}
\label{fig:uml:reporting}
\end{figure}

\begin{figure}
\includegraphics[angle=90,width=\textwidth, keepaspectratio=true]{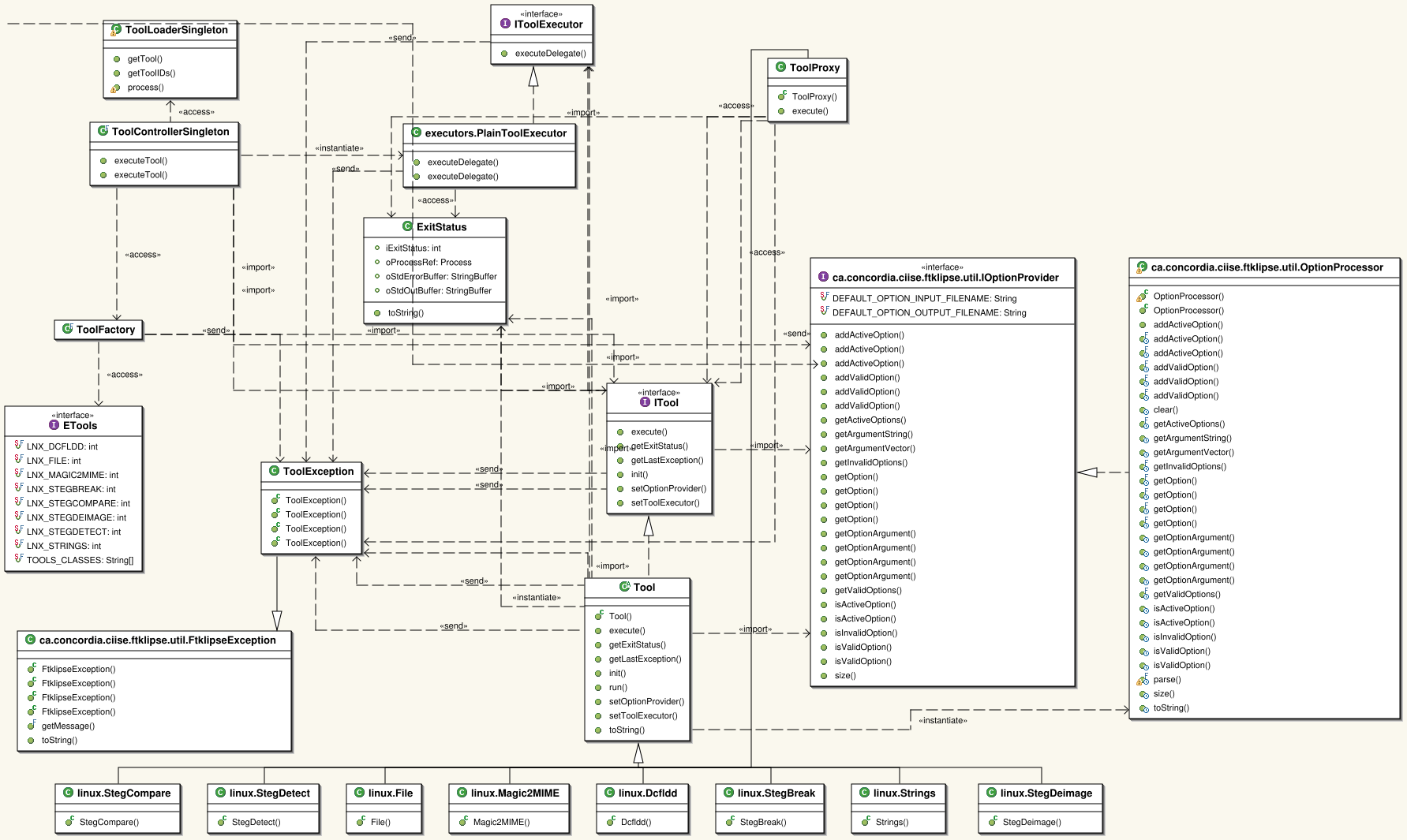}
\caption{Class Diagram for the Backend Tools Framework}
\label{fig:uml:tools}
\end{figure}

\section{Data Storage Format}
\index{Design!Data Storage Format}
\index{Data Storage Format}

This section is about data storage issues and the details
on the chosen undelying implementation and ways of addressing
those issues.

\subsection{Entity Relationship Diagram}
\index{Data Storage Format!ER Diagram}

The ER diagram of the underlying SQL engine we chose
is in \xf{fig:er}. The database is pretty simple as
the \texttt{case\_data} field is a BLOB to which the
\api{Case} data structure is serialized. The \texttt{id\_count}
table is simply there to contain the maximum ID used accros
the database objects in the application. It is updated on
application close, so when the application is loaded back
again, it sets its internal ID from the database properly
for newly created cases and other objects.

\begin{figure}
\includegraphics[width=\textwidth,keepaspectratio=true]{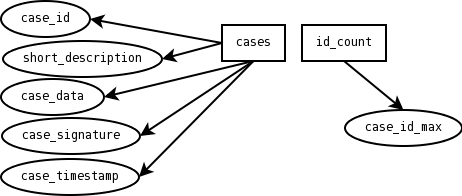}
\caption{Simple ER Diagram of the Internal Database}
\label{fig:er}
\end{figure}

The database is slatted for extension with some code map
data for the UI as well as log facilities later on for
better reporting, like who, what, when, etc.

\subsection{External Systems and Databases}
\index{Data Storage Format!External Systems and Databases}

The database engine the Ftklispe application talks to is abstracted
away so that the actual engine particularities (e.g. SQL queries or XML
atoms) are not visible to
the application thus making it database-engine independent. The provision
was made to have SQL, XML, JavaSpaces\index{Libraries!JavaSpaces} \cite{javaspaces}, or raw object serialization
databases. The actual external database engine used in the demo version of
the toolkit, is the HSQLDB \cite{hsqldb}\index{Libraries!HSQLDB} database, which is implemented
in {\java} itself and has an in-process execution capability.
This database engine is started automatically within the same process
as an application when a first connection is made. It is shudown when application exits.
This choice is justified by simplicity and does not require an external database server to
be set up. This external implementation of the engine is in \file{lib/hsqldb.jar}.

The database-produced files are stored in the \file{data} directory relative
to the current execution environment. The files are \file{ftklipsedb.properties}
and \file{ftklipsedb.script}. The former describes the global database settings
and the latter is the serilized database itself, including DDL
DML statements to reproduce the database. Both are managed by the HSQLDB engine
itself. Originally when deploying the application, neither may present. They
will be created if not present when Ftklipse starts.

Another external system we rely on in the form of library is
the PDF generation library iText \cite{itext}\index{Libraries!iText} \cite{itext}, which is in \file{lib/itext.jar}.
This library is used in \api{PDFReportGenerator} to produce
a PDF copy of the case data stored in the database.

\subsection{Log File Format}
\index{Data Storage Format!Log File Format}

The log is saved in the \file{ftklipse.application.log}. As of this version, the
file is produced with the help of the \api{Logger} class that has been imported
from {\marf} \cite{marf}. (Another logging facility that was considered but not yet
implemented is the Log4J tool \cite{log4j}, which has a full-fledged logging engine.)
The log file produced by \api{Logger} has a classical format of \texttt{[ time stamp ]: message}.
The logger intercepts all attempts to write to STDOUT or STDERR and makes
a copy of them to the file.

\section{Directory and Package Organization}
\index{Directory Structure}
\index{Package Organization}

In this section, we introduce the reader to the structure of the folders
for ftklipse. Please note that Java, by default, converts sub-packages
into subfolders, which is what we see in \xf{fig:folders}.

Please also refer to \xt{tab:folders:explanation} and \xt{tab:packages} for description of
the data contained in the folders and the package organization, respectively.

\begin{figure}[hb]
  \includegraphics[height=\textheight]{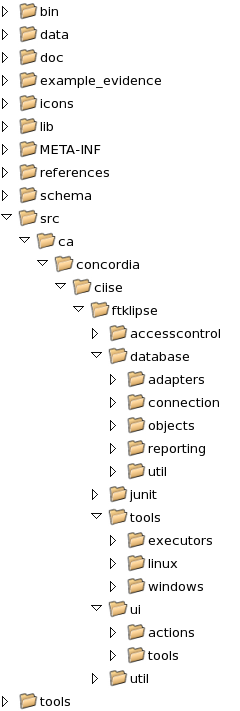}
  \caption{Folder Structure of the Project}
  \label{fig:folders}
\end{figure}

\begin{table}
\begin{tabular}{|l|p{3.5in}|}
	\hline
	\textbf{Folder}				& \textbf{Description} \\ \hline
	bin					& Directory containing the compiled files. All package names described
						  here are also present under this directory. \\ \hline
	data					& Directory containing the case database as well as subdirectories for each of the cases.\\ \hline
	doc					& Project's documentation \\ \hline
	example\_evidence			& Demo evidence that can be used in the projects \\ \hline
	icons					& icons useable for branding and decorating the application \\ \hline
	lib					& External libraries used by ftklipse \\ \hline
	META-INF				& Project's meta-information that would be included in a JAR bundle \\ \hline
	references				& Some useful references on the web on Eclipse development \\ \hline
	schema					& Project's extension point definitions \\ \hline
	src					& Directory containing the source code files. All package names described
						  here are underneath this directory \\ \hline
	tools					& Precompiled tools to use. Also organized hierarchically. \\ \hline

\end{tabular}
\caption{Details on folder structure}
\label{tab:folders:explanation}
\end{table}

\begin{table}
\begin{tabular}{|l|p{1.5in}|}
	\hline
	\textbf{Package}			& \textbf{Description}  \\ \hline
	ca.concordia.ciise.ftklipse		& Ftklipse's root package name \\ \hline
	ca.concordia.ciise.ftklipse.accesscontrol	& Ftklipse's access control model \\ \hline
	ca.concordia.ciise.ftklipse.database		& Ftklipse's database module \\ \hline
	ca.concordia.ciise.ftklipse.database.adapters		& Database adapters \\ \hline
	ca.concordia.ciise.ftklipse.database.connection		& Database connection objects \\ \hline
	ca.concordia.ciise.ftklipse.database.objects		& Object model that is saved and restored from the database \\ \hline
	ca.concordia.ciise.ftklipse.database.reporting		& Reporting sub-module \\ \hline
	ca.concordia.ciise.ftklipse.database.util		& Database utility classes \\ \hline
	ca.concordia.ciise.ftklipse.junit		& Some JUnit tests \\ \hline
	ca.concordia.ciise.ftklipse.tools		& Tool execution module, not including GUI screens \\ \hline
	ca.concordia.ciise.ftklipse.tools.executors		& Tool execution adapters for the underlying platform \\ \hline
	ca.concordia.ciise.ftklipse.tools.linux			& Tool adapters for Linux tools \\ \hline
	ca.concordia.ciise.ftklipse.tools.windows		& Tool adapters for Windows tools \\ \hline
	ca.concordia.ciise.ftklipse.ui			& Ftklipse's user interface classes\\ \hline
	ca.concordia.ciise.ftklipse.ui.actions			& Eclise actions for the menu and right-click menu\\ \hline
	ca.concordia.ciise.ftklipse.ui.tools			& User interfaces for the tools provided by default\\ \hline
	ca.concordia.ciise.ftklipse.util		& Utility classes \\ \hline
\end{tabular}
\caption{Package organization}
\label{tab:packages}
\end{table}

\section{Plug-Ins}
In order to allow tools to be plugged in, we use Eclipse's default mechanism, which requires
to define and export and extension point. The extension point \xt{tab:extension_point} defines a set of properties that
are mostly used to populate the user interface as well as providing the interfaces that must be implemented
in order to contribute a plug-in to ftklipse.

\begin{table}

\begin{tabular}{|l|l|p{2.5in}|}
	\hline
	\textbf{Attribute} 	& \textbf{Type} & \textbf{Summary}  \\ \hline
	id			& string	& unique identifier for the tool \\ \hline
	name			& string	& name of the tool. Not currently used \\ \hline
	class			& ITool		& class implementing our standard interface for the tool execution \\ \hline
	type			& enumeration	& one of collection, analysis or other. Used for structuring tools in menus \\ \hline
	parameter		& string	& for future use, allowing a tool to register more than once but with different paramters that would let it act differently. \\ \hline
	outputfile		& string	& for future use, allowing a tool to register and specify a default output file for its operation \\ \hline
	category		& string	& for future use, in order to group tools for batch collection or batch analysis of data \\ \hline
	platform		& enumeration	& either win or unix. To specify on which platform the tool operates \\ \hline
	inBatchMenu		& boolean	& whether the plug-in requires to be registered in batch processing menus \\ \hline
	inRightClickMenu	& boolean	& whether the plug-in requires to be registered in the right-click menu \\ \hline
	friendlyName		& string	& short name of the tool, for displaying the user \\ \hline
	uiclass			& ITooUI	& class implementing our standard interface for the tool execution \\ \hline

\end{tabular}
\label{tab:extension_point}
\caption{Extension Point for Third-Party Plug-Ins}
\end{table}

Any third party can contribute a plug-in tool in ftklipse by creating an Eclipse plug-in project that
chooses to extend \texttt{ca.concordia.ciise. ftklipse.ftklipse\_tools}. Those plug-ins can afterwards be
installed manually in the Eclipse folder's sub-root, or using Eclipse's built-in installer and updater.
When installed properly, ftklipse will detect them without the need to update any configuration file
or perform other similar adminsitrative works.

Each plug-in is responsible for implementing its own dialog(s) and may optionally define its own parameters persistence
mechanism, although our API strongly sugests the use of Eclipse's technology to do so.

In order that all tools can have access to information from the user interface, and that the user interface
can have access to information about all tools, we used a set of registry singletons which are responsible
to conserve single instances of the information.

Plug-in developpers would thus find the \texttt{WidgetRegistrySingleton} to be very helpful,
as it notably returns a reference to the case and evidence tree, which can be queried
to find the active evidence and active projects.

As such, we do not implement a strict Model-View-Controller (MVC) architecture,
but merely a model that is similar to it, as the plug-ins are trusted not to modify
and user interface elements.

\section{User Interface Design}
\subsection{Appearance}
Ftklipse is implemented using JFace and SWT, technologies provided within the Eclipse framework.
It consists of a single window composed of a menu bar on the top, a tree structure on the left-hand side,
and a multiply-tabbed area at the centre.

This central area displays information about the currently opened evidence file or case information from
the case database. Please refer to \xf{fig:show:intro} and \xf{fig:show:evidence} for screenshots of the implementation.

\begin{figure}
\includegraphics[scale=0.4, angle=90, keepaspectratio=true]{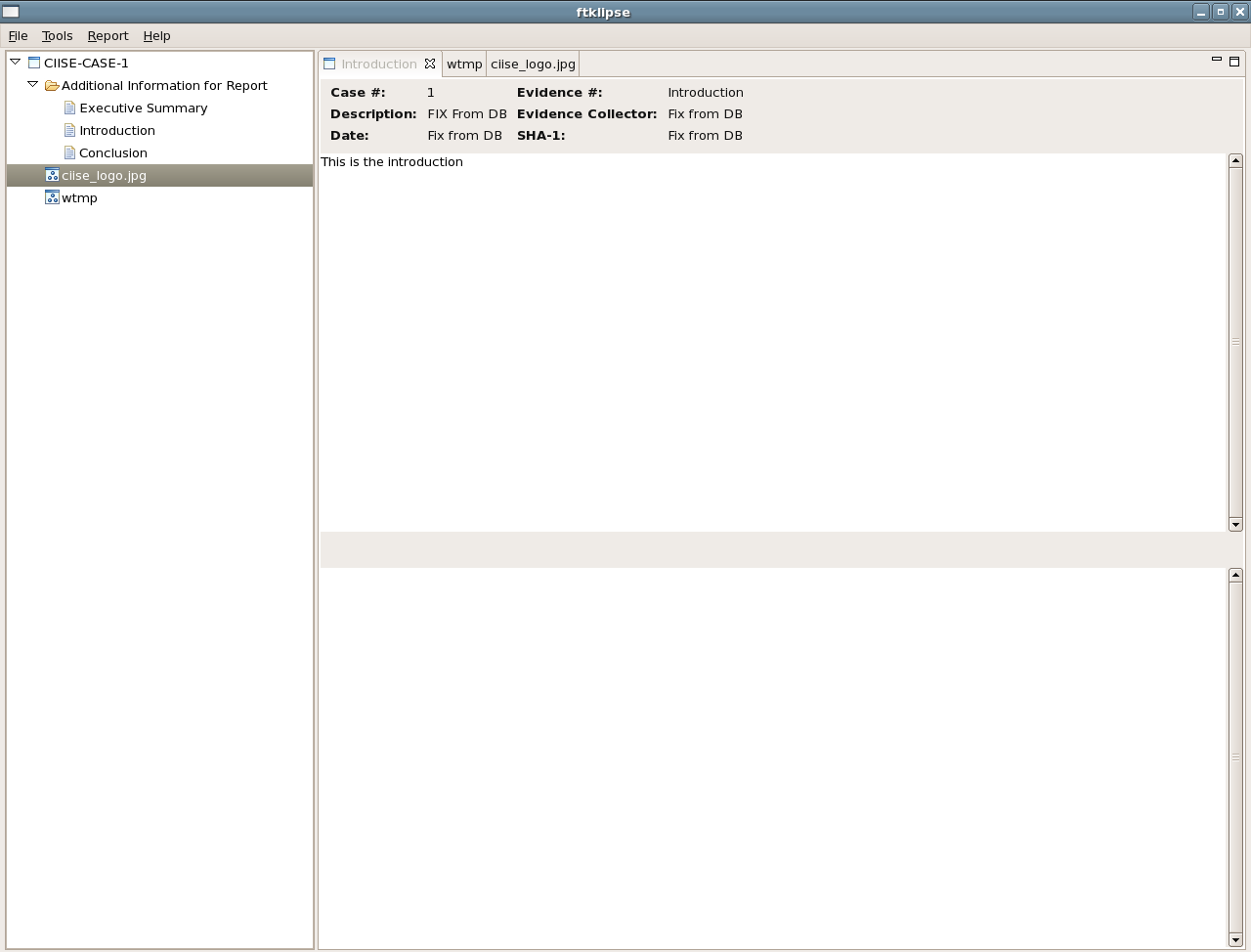}
\caption{User Interface Showing the Case Introduction}
\label{fig:show:intro}
\end{figure}

\begin{figure}
\includegraphics[scale=0.4, angle=90, keepaspectratio=true]{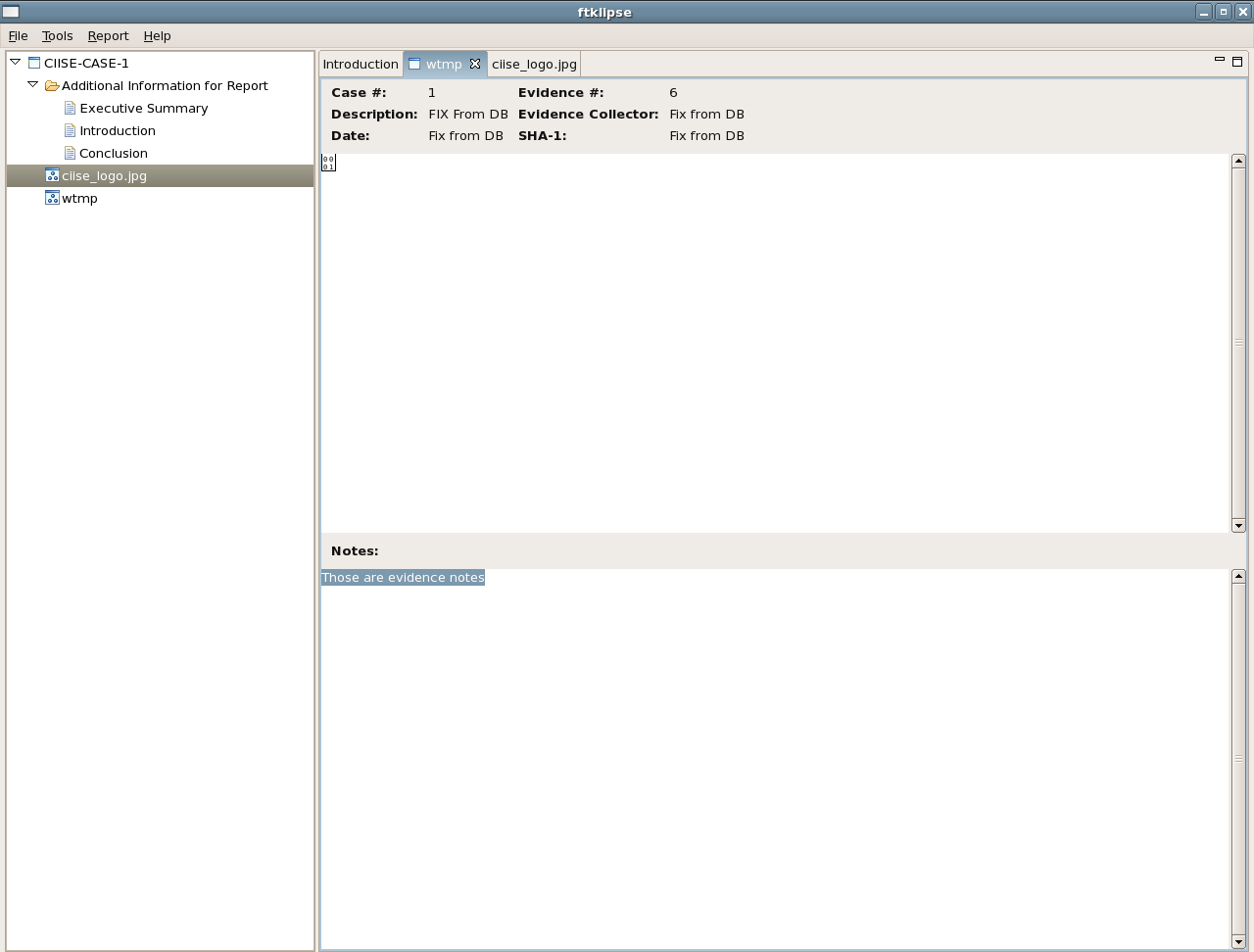}
\caption{User Interface Showing the Evidence Information and Notes}
\label{fig:show:evidence}
\end{figure}

%
%\subsection{Data Model}

%{\todo}

% EOF

%% file: conclusion.tex
\chapter{Conclusion}

Despite the technological difficulties and limitations
the chosen approach seems very promising. Highly modular
design allows also swapping module implementaions from one
technology to another if need be making it very extensible.
Case management, very strong backend architecture for Tools, Database, and Report Generation.
Eclipse UI integration are strong points of this project.

\section{Summary of Technologies Used}

The following were the most prominent technologies used throughout the project:

\begin{itemize}
\item Eclipse IDE\cite{eclipse}
\item iText PDF generation library \cite{itext}
\item HSQLDB lighweight embedded Java SQL engine \cite{hsqldb}
\item Visual Editor for Eclipse \cite{eclipse-visual-editor}
\end{itemize}

\section{Summary of Tools Added}

The number of testing tools is not large and many more
could be added from various resources \cite{forensics-tasks},
however, there were enough for many test cases given time limitations.
The following Linux tools were used for testing and worked:

\begin{itemize}
\item
\tool{stegdetect} \cite{stegdetect}, \tool{stegbreak}, \tool{stegdeimage},
\tool{magic2mime},

\item
\tool{file},

\item
\tool{strings},

\item
\tool{dcfldd}.
\end{itemize}

\section{Summary of Difficulties}

Learning curve for Eclipse plug-in and UI frameworks \cite{eclipse-plugins,eclipse-plugin-forums,eclipse-plugin-dev,eclipse-plugin-i18n,eclipse-plugins-exposed,eclipse-rcp-tutorial,eclipse-plugin-workbench,eclipse-plugin-rcp,eclipse-plugin-extension-points}
with large volumes of APIs and documentation was overwhelming at the beginning
and making things like right- and double-click to work as well as SWT-based \cite{swt}.
UIs was sometimes non-trivial.

\section{Limitations and Technological Restrictions}

The Eclipse framework imposes some technological restrictions in user interface programming on two major areas
that impacted our design.

The first restriction is that the menu items are populated by `Actions', and that it is impractical to have
a different Action instance for each menu item for each possible item the menu can interact with. For example,
the right-click menu, although capable of being dynamically generated every time, requires to perform an action
based on the currently selected item. Re-creating  the menu on each right-click from new objects is expensive
both in memory and computationally, risking to create an interface with a high response time to the user, which impacts
negatively on usability. Another option is to create a cache of such items and change internal data members related
to the selected widget before displaying the menu. This approach increases complexity and was not considered to be a
good solution in our context, due to the complexity of propagating this strategy to existing and future options. Finally,
we considered having a central access point to the information on the selected items that would be opaque to the underlying
data types creating the tree hierarchy. This last approach, altough less `pure' object-oriented design, was retained
for its ease of use in prototyping new features, as well as the assumed atomicity of GUI operation (i.e. it should not be
possible to change the selection while the handling of the right-click on the selection is running).

The second restriction is Eclipse's all-or-nothing approach to plug-in development. As far as we understood the framework,
it is possible to use Eclipse's internal data types and existing advanced widgets only when extending the framework in our plug-in.
A plug-in that would choose not to follow Eclipse's organization (which is our case) could thus not have access to pre-existing
file browsers and variety of editors. As such, the tree hierarchy, mouse handling, and data visualization needed to be reimplemented
from lower-level SWT components.

\section{Future Work and Work-In-Progress}

Allow addition of tools dynamically though GUI
Improve case management with full chain of custody (backend is done for this)
Integration of the hexadecimal editor plugin \cite{ehep}

\section{Acknowledgments}

\begin{itemize}
\item Dr. Mourad Debbabi for the excellent course.
\item Open Source community for Eclipse, HSQLDB, iText
\item Dr. Peter Grogono for {\LaTeX} introductory tutorial \cite{grogono2001}
\end{itemize}

% EOF

%% file: references.tex
\addcontentsline{toc}{chapter}{References}

% 1.Software links for forensics investigative Tasks
% http://www.dmares.com/maresware/SITES/tasks.htm
%http://standards.ieee.org/reading/ieee/std_public/description/se/830-1993_desc.html
%http://standards.ieee.org/reading/ieee/std_public/description/se/830-1998_desc.html

% Look for sdd.bib file if you seek bibliography entries.
\bibliography{sdd}
\bibliographystyle{alpha}

% EOF

%% file: appendix.tex
%Appendix
%\chapter{Something Appendiceful}

%\chapter{ChangeLog}

%\vspace{15pt}
%\hrule
%{\scriptsize \input{ChangLog}}
%\hrule
%\vspace{15pt}

% EOF

%% file: sdd.bbl
\newcommand{\etalchar}[1]{$^{#1}$}
\begin{thebibliography}{{Con}06d}

\bibitem[AGS{\etalchar{+}}06]{log4j}
N.~Asokan, Ceki Gulcu, Michael Steiner, {IBM Zurich Research Laboratory}, and
  {OSS Contributors}.
\newblock {\em log4j, Hierachical Logging Service for Java}.
\newblock apache.org, 2006.
\newblock \url{http://logging.apache.org/log4j/}.

\bibitem[Bol03]{eclipse-plugins}
Azad Bolour.
\newblock {\em Notes on the Eclipse Plug-in Architecture}.
\newblock eclipse.org, July 2003.
\newblock
  \url{http://www.eclipse.org/articles/Article-Plug-in-architecture/plugin_arc%
hitecture.html}.

\bibitem[Bur06]{eclipse-rcp-tutorial}
Ed~Burnette.
\newblock {\em Rich Client Tutorial}.
\newblock eclipse.org, February 2006.
\newblock \url{http://www.eclipse.org/articles/Article-RCP-1/tutorial1.html}.

\bibitem[Col07]{svn}
Inc. CollabNet.
\newblock {\em Subversion (SVN)}.
\newblock tigris.org, 2007.
\newblock \url{http://subversion.tigris.org/}.

\bibitem[{Con}06a]{eclipse-plugin-extension-points}
{Contributors}.
\newblock {\em Creating and using Extension Points}.
\newblock refractions.net, 2006.
\newblock
  \url{http://udig.refractions.net/confluence/display/DEV/1+Creating+and+Using%
+Extension+Points}.

\bibitem[{Con}06b]{eclipse-plugin-forums}
{Contributors}.
\newblock {\em Eclipse Plugin Central - Forums}.
\newblock eclipseplugincentral.com, 2006.
\newblock
  \url{http://www.eclipseplugincentral.com/PNphpBB2+file-viewforum-f-74.html}.

\bibitem[{Con}06c]{swt}
{Contributors}.
\newblock {\em SWT: The Standard Widget Toolkit}.
\newblock eclipse.org, 2006.
\newblock \url{http://www.eclipse.org/swt/}.

\bibitem[{Con}06d]{eclipse-plugin-rcp}
{Contributors}.
\newblock {\em User Guide: Building a Rich Client Platform application}.
\newblock eclipse.org, 2006.
\newblock
  \url{http://help.eclipse.org/help31/index.jsp?topic=/org.eclipse.platform.do%
c.isv/guide/rcp.htm}.

\bibitem[{Con}06e]{eclipse-visual-editor}
{Contributors}.
\newblock {\em Visual Editor Project}.
\newblock eclipse.org, 2006.
\newblock \url{http://wiki.eclipse.org/index.php/Visual_Editor_Project}.

\bibitem[{Con}06f]{eclipse-plugin-workbench}
{Contributors}.
\newblock {\em Workbench User Guide: Plugging into the workbench}.
\newblock eclipse.org, 2006.
\newblock
  \url{http://help.eclipse.org/help31/index.jsp?topic=/org.eclipse.platform.do%
c.isv/guide/workbench.htm}.

\bibitem[E{\etalchar{+}}08]{eclipse}
{Eclipse contributors} et~al.
\newblock {Eclipse Platform}.
\newblock eclipse.org, 2000-2008.
\newblock \url{http://www.eclipse.org}, last viewed April 2008.

\bibitem[Gal02]{eclipse-plugin-dev}
David Gallardo.
\newblock {\em Developing Eclipse plug-ins}.
\newblock ibm.com, December 2002.
\newblock
  \url{http://www-128.ibm.com/developerworks/opensource/library/os-ecplug/?Ope%
n&ca=daw-ec-dr}.

\bibitem[Gro01]{grogono2001}
Peter Grogono.
\newblock {\em A {\LaTeX2e} Gallimaufry. Techniques, Tips, and Traps}.
\newblock Department of Computer Science and Software Engineering, Concordia
  University, Montreal, Canada, March 2001.
\newblock \url{http://www.cse.concordia.ca/~grogono/Writings/gallimaufry.pdf},
  last viewed May 2008.

\bibitem[htt06]{forensics-tasks}
http://www.dmares.com.
\newblock {\em {Software Links for Forensics Investigative Tasks}}.
\newblock 2006.
\newblock \url{http://www.dmares.com/maresware/SITES/tasks.htm}.

\bibitem[KFL02]{eclipse-plugin-i18n}
Dan Kehn, Scott Fairbrother, and Cam-Thu Le.
\newblock {\em Internationalizing your Eclipse plug-in}.
\newblock ibm.com, June 2002.
\newblock
  \url{http://www-128.ibm.com/developerworks/opensource/library/os-i18n/}.

\bibitem[LS06]{itext}
Bruno Lowagie and Paulo Soares.
\newblock {\em iText, a Free Java-PDF library}.
\newblock lowagie.com, 2006.
\newblock \url{http://www.lowagie.com/iText/}.

\bibitem[Mam05]{javaspaces}
Qusay~H. Mamoud.
\newblock {\em Getting Started With JavaSpaces Technology: Beyond Conventional
  Distributed Programming Paradigms}.
\newblock Sun Microsystems, Inc., July 2005.
\newblock
  \url{http://java.sun.com/developer/technicalArticles/tools/JavaSpaces/}.

\bibitem[Pal06]{ehep}
Marcel Palko.
\newblock {\em Eclipse Hex Editor Plugin}.
\newblock sourceforge.net, 2006.
\newblock \url{http://ehep.sourceforge.net/}.

\bibitem[Pro04]{stegdetect}
Niels Provos.
\newblock Steganography detection with stegdetect, 2004.
\newblock \url{http://www.outguess.org/detection.php}.

\bibitem[Pro05]{eclipse-plugins-exposed}
Emmanuel Proulx.
\newblock {\em Eclipse Plugins Exposed}.
\newblock onjava.com, February 2005.
\newblock \url{http://www.onjava.com/pub/a/onjava/2005/02/09/eclipse.html}.

\bibitem[{The}08]{hsqldb}
{The hsqldb Development Group}.
\newblock {HSQLDB} -- lightweight 100\% {Java SQL} database engine v.1.8.0.10.
\newblock hsqldb.org, 2001--2008.
\newblock \url{http://hsqldb.org/}.

\bibitem[{The}09]{marf}
{The MARF Research and Development Group}.
\newblock {The Modular Audio Recognition Framework and its Applications}.
\newblock SourceForge.net, 2002--2009.
\newblock \url{http://marf.sf.net}, last viewed December 2008.

\end{thebibliography}
